\documentclass[a4paper]{jpconf}
\usepackage{graphicx}
\usepackage{subfigure}

\begin{document}
\title{Spectra and elliptic flow for $\Lambda$, $\Xi$,
and $\Omega$ in 200 A GeV Au+Au collisions}

\author{Xiangrong Zhu$^{1,2}$ and Huichao Song$^{1, 2, 3}$}
\address{$^{1}$Department of Physics and State Key Laboratory of Nuclear Physics and Technology, Peking
University, Beijing 100871, China }
\address{$^{2}$Collaborative Innovation Center of Quantum Matter, Beijing 100871, China}
\address{$^{3}$Center for High Energy Physics, Peking University, Beijing
100871, China}

\begin{abstract}
Using {\tt VISHNU} hybrid model, we calculate the $p_{\rm T}$-spectra and elliptic flow of $\Lambda$, $\Xi$,
and $\Omega$ in 200 A GeV Au+Au collisions. Comparisons with the STAR measurements
show that the model generally describes these soft hadron data. We also briefly study and discuss the mass ordering of elliptic flow among $\pi$, $K$, $p$, $\Lambda$, $\Xi$, and $\Omega$ in minimum bias Au+Au collisions.

\end{abstract}

\section{Introduction\label{sec:intro}}
\quad The main goal of relativistic heavy-ion collisions is to create the Quark-gluon Plasma (QGP) and study its properties. Due to their small hadronic cross sections, multi-strange hadrons, such as $\Xi$ and $\Omega$,
experience early chemical and thermal freeze-out after hadronization~\cite{vanHecke:1998yu,SongMultistrange}.  Their anisotropy flow are mainly developed in the QGP phase and less contaminated by the hadronic evolution, which are expected to provide valuable information for the QGP.
In our early paper~\cite{SongMultistrange},  the spectra and elliptic flow of strange and multi-strange hadrons at the LHC have been studied with {\tt VISHNU} hybrid model. This proceeding will extend the calculations to top RHIC energies and compare the results with recent STAR data.

\section{Setup of the calculations\label{sec:setup}}
\quad {\tt VISHNU} hybrid model~\cite{Song:2010aq} combines (2+1)-d relativistic viscous
hydrodynamics ({\tt VISH2+1})~\cite{Song:2007fn} for the QGP fluid expansion with a microscopic hadronic transport model({\tt UrQMD})~\cite{Bass:1998ca}
for the hadron resonance gas evolution. The hydrodynamic simulations input an equation of state (EoS) s95p-PCE constructed from recent lattice QCD data~\cite{Huovinen:2009yb}. The transition from hydrodynamics to the hadron cascade is controlled by a switching temperature which is
set at 165 MeV. We input smooth initial conditions that are generated from MC-Glauber model through averaging a large number of fluctuating initial entropy density profiles (Here, each initial density distribution is recenterd and rotated to align the major and minor axes before averaging, which is called as {\it initialization in the participant plane})~\cite{Hirano:2009ah}. Following~\cite{Song:2011hk,SongQM}, we set the initial time $\tau_0=0.6~{\rm fm}/c$, the specific shear viscosity $(\eta/s)_{QGP}=0.08$ and specific bulk viscosity $(\zeta/s)_{QGP}=0$. The normalization factor of the the initial entropy density profiles is tuned to reproduce the final charged hadron multiplicity in the most central Au+Au collisions at $\sqrt{s_{NN}}$=200 GeV~\cite{Abelev:2008ab}. These inputs and settings once nicely described the $p_T$ spectra and differential elliptic flow for all charged hadron and for pions and protons in 200 A GeV Au+Au collisions.

\begin{figure}[tbph]
\centering
  \includegraphics[width=0.9\linewidth,height=9cm,clip=]{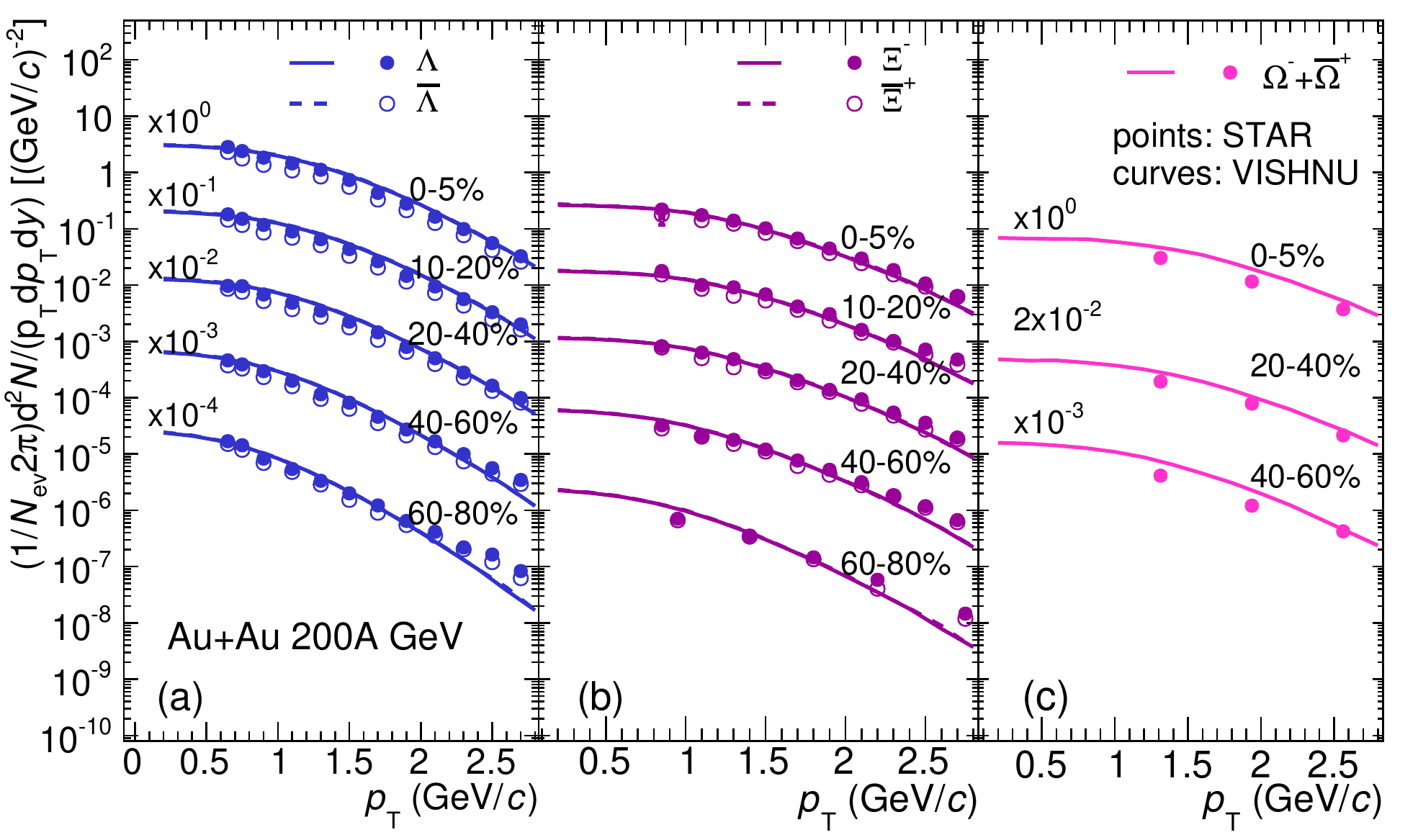}
  \caption{(Color online) Transverse momentum spectra of $\Lambda$, $\Xi$ and $\Omega$ at various centralities in 200 A Au+Au collisions. Experimental data are from the STAR measurements~\cite{Abelev:2008ae}. Theoretical curves are calculated from {\tt VISHNU} with MC-Glauber initial conditions, $\tau_0=0.6 \ \mathrm{fm/c}$ and $\eta/s=0.08$.
 \label{fig:PtLXO}}
\end{figure}

\section{Results}
\quad The $p_{\rm T}$-spectra and elliptic flow of pions and protons in 200 A Au+Au collisions have been
studied with {\tt VISHNU} hybrid model in our early paper~\cite{Song:2010aq}. We showed that, with MC-Glauber initial conditions, $\tau_0=0.6 \ \mathrm{fm/c}$ and $\eta/s=0.08$, {\tt VISHNU} nicely describes these low $p_T$ data of pions and protons at various centrality bins.

Figure~\ref{fig:PtLXO} presents transverse momentum spectra of $\Lambda$, $\Xi$ and $\Omega$ in 200 A GeV Au+Au collisions~\footnote{For the STAR measurement, the $\Lambda$ spectra has been corrected with the feed-down contributions from weak decays. $\Lambda$, $\Xi$ and $\Omega$ from {\tt VISHNU} are  final hadrons after collisions and strong resonance decays since the {\tt UrQMD} hadronic evolution does not contain any weak decay channels.}. In general, {\tt VISHNU} describes the $p_{\rm T}$-spectra of $\Lambda$ and $\Xi$, but slightly over-predicts the production of $\Omega$ for all centrality bins. In spite of the normalization issues, {\tt VISHNU} nicely fits the slope of the $p_{\rm T}$-spectra for $\Lambda$, $\Xi$ and $\Omega$ at various centralities.
Together with the early nice descriptions of the $p_{\rm T}$-spectra for pions, kaons and protons~\cite{Song:2011hk}, it indicates that {\tt VISHNU} generates a proper amount of radial flow during the QGP and hadronic evolution at the top RHIC energy.

\begin{figure}[tbph]
\centering
 \includegraphics[width=0.9\linewidth]{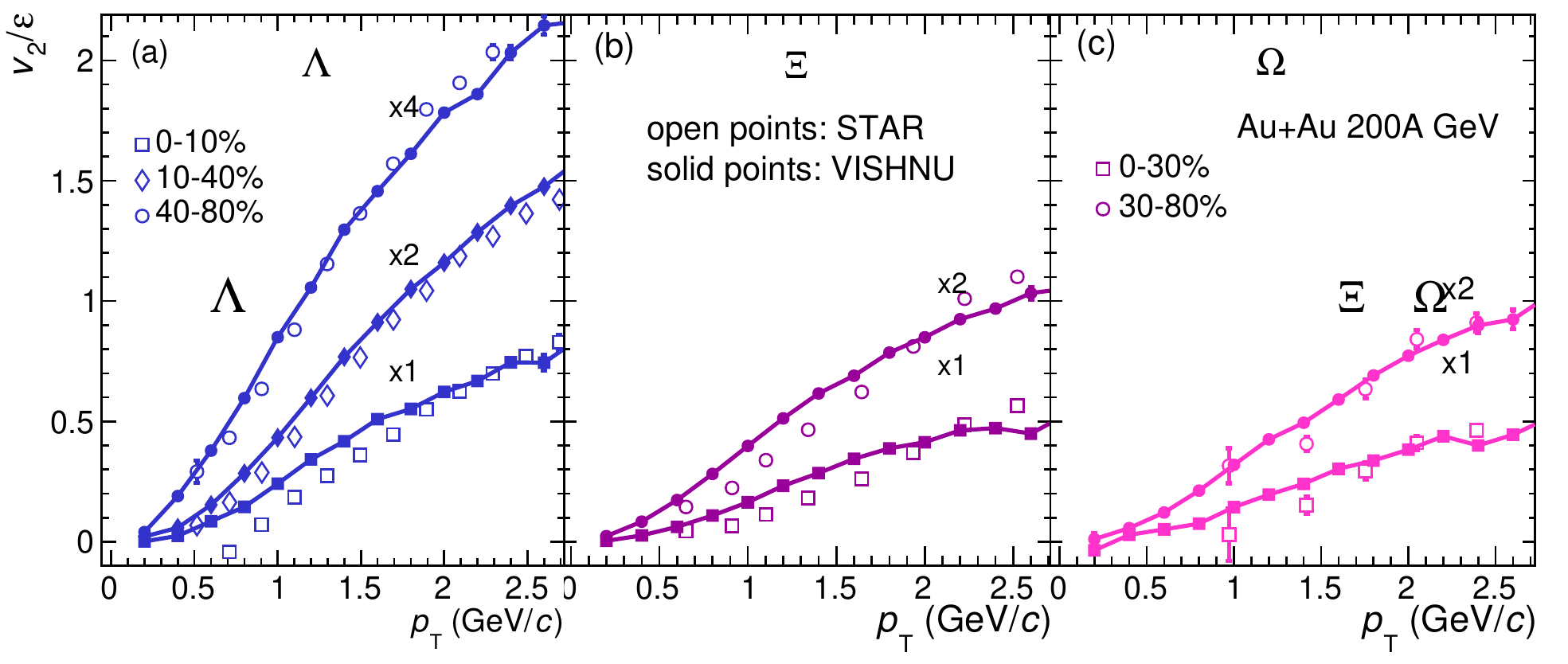}
  \caption{(Color online) Differential elliptic flow of $\Lambda$, $\Xi$ and $\Omega$ in 200 A GeV Au+Au collisions.
  Experimental data are from STAR~\cite{Abelev:2008ae,Adamczyk:2015ukd}, theoretical curves are calculated from {\tt VISHNU} with
  the same inputs as for Fig.~\ref{fig:PtLXO}.
 \label{fig:V2LXO}}
\end{figure}
Figure~\ref{fig:V2LXO} shows the differential elliptic flow of $\Lambda$, $\Xi$ and $\Omega$
in 200 A GeV Au+Au collisions. The theoretical curves are calculated from {\tt VISHNU} using smooth initial conditions from MC-Glauber (with an average in participant plane). The STAR data are measured by the event plane method~\cite{Abelev:2008ae,Adamczyk:2015ukd}, which contain the contributions from event-by-event fluctuations and the non-flow effects. To account the fluctuation contributions, we divide the flow data by $\langle\varepsilon^{\alpha}_{\rm part}\rangle^{\alpha}$ and the theoretical results by $\bar{\varepsilon}_{\rm part}$, where the exponent $\alpha$ depends on the event-plane resolution $R$ and details of the $v_{2}$ extraction method~\cite{Song:2011hk,Ollitrault:2009ie}. The detail values of $\langle\varepsilon^{\alpha}_{\rm part}\rangle^{\alpha}$ and $\bar{\varepsilon}_{\rm part}$ can be found in~\cite{Song:2011hk}. Our past research has shown that {\tt VISHNU} nicely fits the elliptic flow for all charged hadrons and for identified hadrons $\pi$ and $p$ in 200 A GeV Au+Au collisions with MC-Glauber, $\tau_0=0.6 \ \mathrm{fm/c}$ and  $\eta/s=0.08$~\cite{Song:2011hk}. With the same inputs, we extend the {\tt VISHNU} calculations for the strange and multi-strange hadrons with high statistical simulations. Fig.~2 shows that, {\tt VISHNU} generally fits the elliptic flow of $\Lambda$, $\Xi$ and $\Omega$ in 200 A GeV Au+Au collisions. For 0-10\% centrality, the {\tt VISHNU} results are slightly above the data for all three hadrons. It is worthwhile to mention that, for 2.76 A TeV Pb+Pb collisions, early {\tt VISHNU} calculations also slightly over-predict the elliptic flow of $\Lambda$, $\Xi$ and $\Omega$ at 10-20\% and 30-40\% centralities. In fact,  the {\tt UrQMD} simulations in {\tt VISHNU} assigns small cross-sections to  $\Xi$ and $\Omega$, leading to an early decoupling of these two hadrons. Compared with pure hadronic simulations~\cite{Shen:2011eg}, this leads to an insufficient development of elliptic flow for these two multi-strange hadrons in central and semi-central collisions at top RHIC and the LHC energies.

It is generally believed that the mass ordering of elliptic flow among various identified hadrons reflects the interplay between radial and elliptic flows, providing more information on the QGP fireball evolution. Figure 3 presents the mass ordering of differential elliptic flow among $\pi$, $K$, $p$, $\Lambda$, $\Xi$, and $\Omega$ in minimum bias Au+Au collisions. Compared with the STAR measurements, {\tt VISHNU} correctly describes the $v_2$ mass ordering among $\pi$, $K$, $p$, and $\Omega$, but fails to reproduce the mass ordering of $\Lambda$ and $\Xi$.  In our early investigations~\cite{SongMultistrange,Song:2013qma}, a similar situation was found in 2.76 A TeV Pb+Pb collisions at 10-20\% and 40-50\% centralities.  Although {\tt VISHNU} correctly generates the $v_2$ mass ordering among $\pi$, $K$, $p$, it slightly under-predicts the proton $v_2$ at lower transverse momentum. Together with the slightly over-perditions of the $\Lambda$ and $\Xi$ data, this leads to an inverse $v_2$ mass-ordering among $p$, $\Lambda$ and $\Xi$. The effects from the initial flow or improved {\tt UrQMD} hadronic cross-sections may solve this issue within the framework of {\tt VISHNU}, which should be investigated in the near future.

\section{Summary\label{sec:summary}}
\quad In this proceeding, we present {\tt VISHNU} calculations on the soft hadron data of strange and multi-strange hadrons in 200 A GeV Au+Au collisions. With the parameter settings used in our early calculations, {\tt VISHNU} generally describes the $p_{\rm T}$-spectra and elliptic flow of  $\Lambda$, $\Xi$ and $\Omega$ for various centrality bins. We also compare the $v_2$ mass ordering among $\pi$, $K$, $p$, $\Lambda$, $\Xi$, and $\Omega$ with the STAR measurements and find that {\tt VISHNU} correctly describes the $v_2$ mass ordering among $\pi$, $K$, $p$, and $\Omega$, but fails to reproduce the mass ordering among $p$, $\Lambda$, and $\Xi$.

\begin{figure}[t]
\centering
\includegraphics[width=0.9\linewidth]{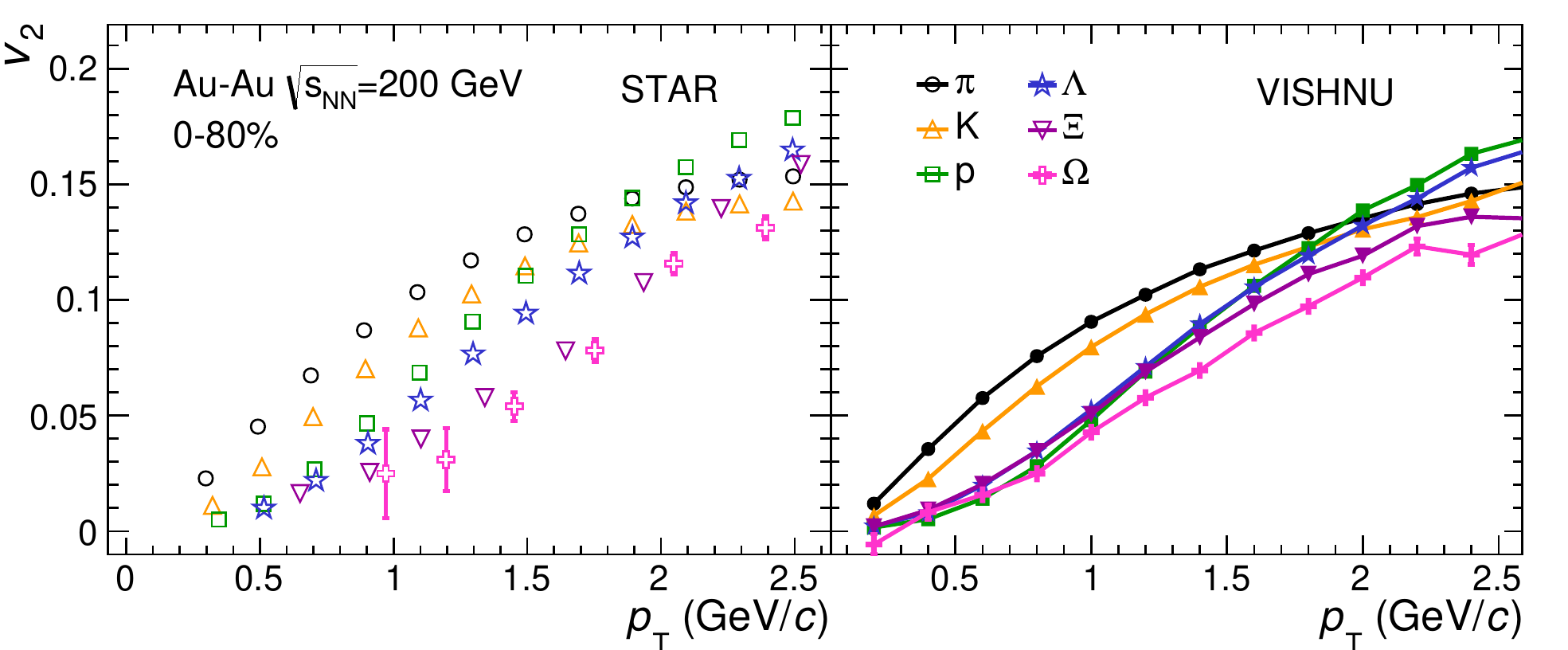}
\caption{(Color online) Differential elliptic flow of $\pi$, $K$, $p$, $\Lambda$, $\Xi$, and $\Omega$
  in minimal bias Au+Au collisions. Left panel: STAR measurements~\cite{Adamczyk:2015ukd,Adams:2006ke},
  right panel: {\tt VISHNU} calculations.
\label{fig:V2mass}}

\end{figure}

\ack
We thank X. Zhu and S. Shi for helpful discussion. This work was supported by the NSFC and the MOST under Grants No. 11435001 and No. 2015CB856900.  X.Z. was also partially supported by the China Postdoctoral Science Foundation under Grant No. 2015M570879. We gratefully acknowledge extensive computing resources provided to us on Tianhe-1A by the National Supercomputing Center in Tianjin, China.


\section*{References}
\begin{thebibliography}{9}

\bibitem{vanHecke:1998yu}
  van Hecke H, Sorge H and Xu N 1998
  {\it Phys.\ Rev.\ Lett.}  {\bf 81} 5764

\bibitem{SongMultistrange}
 Zhu X, Meng F, Song H, and Liu Y X 2015
{\it Phys. Rev.}  C {\bf 91} 034904 ;
Song H, Meng F, Xin X and Liu Y X 2014
  {\it J.\ Phys.\ Conf.\ Ser.}  {\bf 509} 012089



\bibitem{Song:2010aq}
  Song H, Bass S A and Heinz U 2011
  {\it Phys.\ Rev.}  C {\bf 83} 024912;
Song H 2015
  {\it Pramana} {\bf 84} 5 703-15,  [arXiv:1401.0079 [nucl-th]]


\bibitem{Song:2007fn}
  Song H and Heinz U 2008
  {\it Phys.\ Lett.}  B {\bf 658} 279;
  2008 {\it Phys.\ Rev.}  C {\bf 77} 064901;
  2008 {\it Phys.\ Rev.} C {\bf 78} 024902;
  Song H 2009 {\it Ph.D thesis} ({\it Preprint}
  arXiv:0908.3656 [nucl-th])

\bibitem{Bass:1998ca}
  Bass S A {\it et al.} 1998
  {\it Prog.\ Part.\ Nucl.\ Phys.}  {\bf 41} 255;
%
  Bleicher M {\it et al.} 1999
  {\it J.\ Phys.} G {\bf 25} 1859

\bibitem{Huovinen:2009yb}
  Huovinen P and Petreczky P 2010
  {\it Nucl.\ Phys.} A {\bf 837} 26-53


\bibitem{Hirano:2009ah}
  Hirano T and Nara Y 2009
  {\it Phys.\ Rev.}  C {\bf 79} 064904

\bibitem{Song:2011hk}
  Song H, Bass S A, Heinz U, Hirano T and Shen C 2011
  {\it Phys.\ Rev.} C {\bf 83} 054910; 2012 {\it Phys.\ Rev.} C {\bf 86} 059903

\bibitem{SongQM}
Song H 2013
  {\it Nucl.\ Phys.} A {\bf 904-905} 114c-21c, [arXiv:1210.5778 [nucl-th]].


\bibitem{Abelev:2008ab}
  Abelev B I {\it et al.} (STAR Collaboration) 2009
  {\it Phys.\ Rev.} C {\bf 79} 034909


\bibitem{Abelev:2008ae}
  Abelev B I {\it et al.} (STAR Collaboration) 2008
  {\it Phys.\ Rev.} C {\bf 77} 054901

\bibitem{Adamczyk:2015ukd}
  Adamczyk L {\it et al.} (STAR Collaboration) 2015
  ({\it Preprint} arXiv:1507.05247 [nucl-ex])

\bibitem{Ollitrault:2009ie}
  Ollitrault J Y, Poskanzer A M and Voloshin S A 2009
  {\it Phys.\ Rev.} C {\bf 80} 014904

\bibitem{Adams:2006ke}
  Adams J {\it et al.} (STAR Collaboration) 2007
  {\it Phys.\ Rev.\ Lett.}  {\bf 98} 062301

\bibitem{Shen:2011eg}
  Shen C, Heinz U, Huovinen P and Song H 2011
 {\it Phys.\ Rev.}\ C {\bf 84} 044903

\bibitem{Song:2013qma}
  Song H, Bass S and Heinz U W 2014
  {\it Phys. Rev.} C {\bf 89} 034919

\end{thebibliography}
\end{document}